\documentclass[12pt]{article}

\usepackage{graphicx,  amssymb, array, amsmath, float}

\usepackage{color}
\usepackage[authoryear]{natbib}
\usepackage{array}
\usepackage{bm}
\usepackage{graphicx}
\usepackage{float}
\usepackage{subcaption}
\usepackage{multicol}
\usepackage{xr}
\usepackage{mathtools}
\usepackage{comment}
\usepackage{hyperref}
\usepackage{cleveref}
\usepackage{multirow}
\usepackage{xurl}
\usepackage{soul}
\usepackage{nameref}
\newcommand{\blind}{0}

\begin{document}

\def\spacingset#1{\renewcommand{\baselinestretch}%
{#1}\small\normalsize} \spacingset{1}

\if0\blind
{
\title{\bf Data Privacy Protection and Utility Preservation through Bayesian Data Synthesis: A Case Study on Airbnb Listings }

\author{Shijie Guo \\
    Civil and Environmental Engineering Department, Stanford University \\
    and \\
    Jingchen Hu\hspace{.2cm}\\
    Mathematics and Statistics Department, Vassar College}
\maketitle
} \fi

\if1\blind
{
  \bigskip
  \bigskip
  \bigskip
  \begin{center}
    {\LARGE\bf Data privacy protection and utility preservation through Bayesian data synthesis: a case study on Airbnb listings }
\end{center}
  \medskip
} \fi

\bigskip
\begin{abstract}
\noindent When releasing record-level data containing sensitive information to the public, the data disseminator is responsible for protecting the privacy of every record in the dataset, simultaneously preserving important features of the data for users' analyses. These goals can be achieved by data synthesis, where confidential data are replaced with synthetic data that are simulated based on statistical models estimated on the confidential data. In this paper, we present a data synthesis case study, where synthetic values of price and the number of available days in a sample of the New York Airbnb Open Data are created for privacy protection. One sensitive variable, the number of available days of an Airbnb listing, has a large amount of zero-valued records and also truncated at the two ends. We propose a zero-inflated truncated  Poisson regression model for its synthesis. We utilize a sequential synthesis approach to further synthesize the sensitive price variable. The resulting synthetic data are evaluated for its utility preservation and privacy protection, the latter in the form of disclosure risks. Furthermore, we propose methods to investigate how uncertainties in intruder’s knowledge would influence the identification disclosure risks of the synthetic data. In particular, we explore several realistic scenarios of uncertainties in intruder’s knowledge of available information and evaluate their impacts on the resulting identification disclosure risks.
\end{abstract}

\noindent%
\vfill

\newpage
\spacingset{1.45} 

\section{Introduction}

Airbnb, one of the leading properties rental service providers, has rapidly gained popularity among people in need of lodging services, ever since its founding in 2008. An increasing number of travellers decide to choose Airbnb properties over traditional accommodations, such as hotels and hostels, when traveling \citep{guttentag2018tourists}. By 2020, Airbnb has over seven million listings across the globe in more than 220 countries and regions~\citep{airbnb}. 

However, as Airbnb customers and property hosts are enjoying the convenience it provides, concerns about the safety and security of the listed properties keep rising. \citet{xu2017explore} found a positive correlation between the spatial distribution of Airbnb and the number of property crimes. This result suggests that Airbnb properties may be more prone to property crimes than non-rental properties. Moreover, multiple media reports have shown that some Airbnb guests are victims of crimes including theft, sexual assault, and kidnapping, among other things \citep{fox_news_2015, grind_shifflett_2019, bressington_2019}. Thus, when releasing data about Airbnb listings to the public, the data disseminator, which could be Airbnb itself or other holders of Airbnb data, should prioritize the task of privacy protection of each released Airbnb listing. Proper protection of Airbnb listings' sensitive information would prevent potential perpetrator from taking advantage of the released Airbnb data when committing crimes. 

For the privacy protection purpose of releasing Airbnb listings data, our study considers the method of synthetic data, one of the leading approaches to providing privacy protection for the release of record-level data \citep{Rubin1993synthetic, Little1993synthetic}. To create synthetic data, data disseminators first build Bayesian statistical models and estimate them on the confidential record-level data. Next, synthetic values of sensitive variables are simulated from the posterior predictive distributions of the estimated Bayesian models. Data disseminators can then release the synthetic data to the public, after evaluating their utility (i.e., usefulness) and their privacy protection. We refer interested readers to \citet{Drechsler2011book} for a detailed description of the generation and evaluation of synthetic data for privacy protection. We note that government statistical agencies, such as the U.S. Census Bureau, have been creating synthetic data products for public release for more than a decade by now. For example, the OnTheMap Census product \citep{OnTheMap2008} and the synthetic Longitudinal Business Database \citep{SynLBD2011}, among others. There are also non-Bayesian methods for data synthesis, including classification and regression trees (CART), proposed by \citet{CART2005} and can be implemented using the synthpop R package \citep{synthpopR}.

In our case study, the New York Airbnb Open Dataset is considered as the confidential dataset to be protected \citep{bnbData}. Among the available variables in this confidential dataset, the number of available days an Airbnb property for listing and the average listing price of an Airbnb property are deemed sensitive, and therefore to be synthesized for privacy protection. The number of available days variable, in particular, has many zero-valued records and also has a minimum of 0 days and a maximum of 365 days. To fully capture such features, we develop a Bayesian zero-inflated truncated Poisson regression model to synthesize the number of available days. As for the continuous variable of the average price of a listing, we create a Bayesian linear regression model for its synthesis. In each synthesis model, we utilize three predictors, the room type of an Airbnb listing, the neighborhood the listing is in, and the number of reviews the listing has on Airbnb, all of which are deemed insensitive and therefore kept un-synthesized. We build a sequential synthesis process \citep{SynLBD2011}, where the number of available days is synthesized first, and the average price is synthesized next. In the end, we create multiple synthetic datasets with number of days and average price synthesized while the remaining variables are un-synthesized, commonly known as partially synthetic data \citep{Little1993synthetic}.

Next, we evaluate the utility preservation performance of our simulated synthetic datasets through several data utility measures, including global utility measures such as the propensity score, the cluster analysis, and the empirical CDF, and analysis-specific utility measures such as the confidence interval overlap of quantities of interest \citep{woo2009global, Snoke2018JRSSA}. We also evaluate the privacy protection performance of the partial synthetic datasets through disclosure risk evaluations, where we consider both the identification disclosure risk and the attribute disclosure risk \citep{hu2019bayesian}. Attribute disclosure risk refers to the risk that a data intruder correctly infers the true values of the sensitive variables in the synthetic dataset given access to external databases \citep{reiter2014JPC, hu2014disclosure, hu2019bayesian, mitra2020confidentiality}. Identification disclosure risk occurs when a data intruder, with certain knowledge about records in the dataset, successfully establishes a one-to-one relationship between a confidential record with itself in the synthetic dataset \citep{reiter2009estimating}. 
One important aspect and also an open question in the risk evaluation of synthetic data is the assumptions made about intruder's knowledge. To explore uncertainties in these assumptions, we propose and explore innovative risk evaluation approaches that incorporate such uncertainties. 

Through a series of evaluations, we have found that our partial synthesis of the New York Open Airbnb Data preserves the global utility and the analysis-specific utility of the confidential data at a high level, while the attribute and identification disclosure risks of the synthetic datasets are relatively low. When we introduce uncertainties to the intruder’s knowledge, the more uncertain the intruder’s knowledge is, the lower the identification disclosure risks are, an expected and important result. 

We note that some all-purpose synthesis methods, such as CART \citep{CART2005}, are able to handle truncation and inflation of zero values that exist in the number of available days in the Airbnb sample, because CART takes random samples from the observed values in each leaf (the default settings in the \texttt{synthpop} R package \citep{synthpopR}). However, such a process might result in unacceptable disclosure risks, which we investigate with comparison to our proposed Bayesian synthesis models in the Supplementary Materials. 

We now proceed to introduce the New York Airbnb Open Dataset in our case study.

\subsection{The New York Airbnb Open Dataset}

Our New York Airbnb Open Dataset contains 16 different variables for a total of 48,900 Airbnb listings. We focus on five most relevant variables: neighborhood, room type, number of reviews, number of available days, and the average price. Detailed descriptions of these five variables are in Table \ref{tab:data description}. For the purpose of this study, we take a random sample of $n = 10,000$ records from the entire dataset, ensuring records in our selected dataset have valid entries for the five variables of interest.

\begin{table}[t]
\centering
\begin{tabular}{m{1.1in} m{0.8in} m{3.1in} | >{\centering}m{0.6in}}
\hline
Variable Name  & Type & Description & Sensitive\tabularnewline
\hline
Neighborhood & Categorical & The borough a listing is located in & No\tabularnewline
RoomType & Categorical & A listing's space type & No \tabularnewline
ReviewsCount & Count & The number of reviews a listing received on Airbnb & No\tabularnewline
AvailableDays & Count & The number of days in a year a listing is available for renting on Airbnb, with many $0$s and a maximum value of $365$ & Yes \tabularnewline
Price & Continuous & The average price of a listing for one night & Yes\tabularnewline
\hline
\end{tabular}
\vspace{1mm}
\caption{Descriptions of the five selected variables from the New York Airbnb Open Dataset.}
\label{tab:data description}
\end{table}

Among these five variables, AvailableDays and Price are considered sensitive. Therefore, we build Bayesian models to synthesize these two variables, while using the other variables as predictors. The confidential AvailableDays variable has a large number of records with zero values (its 25\% quantile value is 0). Moreover, it has a minimum of 0 days and a maximum of 365 days. For modeling purpose, we apply the log transformation on the Price variable and use the resulting LogPrice variable for our case study. A table showing key summary statistics for the confidential AvailableDays and Price values is included in the Supplementary Materials for further reading. 

The remainder of this paper is organized as follows. Section \ref{BayesianSyn} presents the two Bayesian synthesis models we propose and our sequential synthesis process. Section \ref{Utility} evaluates the data utility of the synthetic datasets through both global and analysis-specific utility measures. Section \ref{Risk} assesses the attribute and identification disclosure risks of the synthetic datasets under certain assumptions of data intruder's knowledge. Moreover, uncertainties in some of these assumptions are explored and investigated. We end the paper with a few concluding remarks in Section \ref{Conclusion}.

\section{Bayesian synthesis models and sequential synthesis}
\label{BayesianSyn}

We generate partially synthetic datasets for our selected sample of the New York Airbnb Open Dataset through a combination of a zero-inflated truncated Poisson regression model for AvailableDays and a linear regression model for Price. The model for Price is sequentially dependent on that for AvailableDays. In this section, we present the details of the synthesis models in Sections \ref{models:Poisson} and \ref{models:linear}, and describe the sequential synthesis process in Section \ref{models:sequential}.

\subsection{Zero-inflated truncated Poisson regression model for AvailableDays}
\label{models:Poisson}

Given its features of being a count variable and having a significant number of zeros with a minumum of 0 and a maximum of 365, we develop a zero-inflated truncated Poisson regression synthesis model to synthesize the sensitive AvailableDays. AvailableDays is likely to depend on Neighborhood, RoomType, and ReviewsCount, therefore we use three variables as predictors in the zero-inflated truncated Poisson regression model, which can be expressed as follows:
\begin{align}
    Y_{days,i} &\sim q_i\cdot0 + (1-q_i)\cdot \textrm{Poisson}(\lambda_i)^{[0, \,365]}, \label{eq:1}\\
    q_i &\sim \textrm{Bernoulli}(p_i), \label{eq:2}\\
    \log(\lambda_i) &= \alpha_{1}X_{1i} + \alpha_{2}X_{2i} + \alpha_{3} \cdot \log{X_{3i}} + \epsilon_i, \label{eq:3}\\
    \textrm{logit}(p_i) &= \beta_{1}X_{1i} + \beta_{2}X_{2i} + \beta_{3}\cdot \log{X_{3i}}, \label{eq:4}\\
    \epsilon_i &\sim \textrm{Normal}(0, \tau), \label{eq:5}
\end{align}

\noindent where $Y_{days,i}$ is AvailableDays for record $i$, $q_i$ is binary following a Bernoulli distribution with parameter $p_i$, and $\lambda_i$ is the rate for the truncated Poisson distribution if AvailableDays for record $i$ is non-zero in Equation (\ref{eq:1}). The notation $\textrm{Poisson}(\lambda_i)^{[0, \,365]}$ refers to its truncation below at 0 and above at 365. As shown in Equation (\ref{eq:3}), $\log(\lambda_i)$ is modeled given the three predictors, $X_{1i}$ for RoomType, $X_{2i}$ for Neighborhood, and $X_{3i}$ for ReviewsCount, all for record $i$. The logit of the probability $p_i$ for AvailableDays being $0$ for record $i$ is expressed as a linear function with the same three predictors, shown in Equation (\ref{eq:4}). We give a $\textrm{Gamma}(0.001, 0.001)$ prior to the parameter $\tau$, the standard deviation of the normal error term $\epsilon_i$ in Equation (\ref{eq:3}). We give a $\textrm{Normal}(0,1)$ prior to the hyperparameters $\alpha_1,$ $\alpha_2$, $\alpha_3$, $\beta_1$, $\beta_2$, and $\beta_3$ in Equation (\ref{eq:3}) and (\ref{eq:4}). These hyperparameters function in this hierarchical model setup to pool information across different combinations of RoomType, Neighbhorhood, and ReviewsCount.

\subsection{Linear regression model for log Price}
\label{models:linear}

To synthesize the sensitive Price variable, a linear regression model dependent on RoomType, Neighborhood, ReviewsCount, and AvailableDays is used to fit the log of Price in the confidential dataset. This linear regression model can be represented as the following:
\begin{align}
    \log{(Y_{price,i})} &\sim \textrm{Normal}(\mu_i, \sigma_i), \label{eq:6}\\
    \mu_i &= \gamma_{1}X_{1i} + \gamma_{2}X_{2i} + \gamma_{3}\cdot X_{3i} + \gamma_{4}\cdot Y_{days,i}, \label{eq:7}\\
    \sigma_i &\sim t(1,1)^{+} \label{eq:8},
\end{align}

\noindent where $Y_{price,i}$ is the confidential Price for record $i$, $\mu_i$ is the mean for the normal distribution, $\sigma_i$ is the standard deviation of the normal distribution in Equation (\ref{eq:6}). Equation (\ref{eq:7}) shows the expression of mean $\mu_i$, where $X_{1i}, X_{2i}, X_{3i}$ are RoomType, Neighborhood, and ReviewsCount respectively, while $Y_{days,i}$ is the confidential AvailableDays for record $i$. The notation $t(1,1)^{+}$ refers to the $t(1, 1)$ distribution on the positive real line for the standard deviation $\sigma_i$ in Equation (\ref{eq:8}). We give a $\textrm{Normal}(0,2)$ prior to the hyperparameters $\gamma_1,$ $\gamma_2$, $\gamma_3$, and $\gamma_4$ in Equation (\ref{eq:7}). 

Moreover, Equation (\ref{eq:7}) indicates that we sequentially synthesize $Y_{days,i}$ first and $Y_{price,i}$ next. We proceed to describe the details of this sequential synthesis process. 

\subsection{The sequential synthesis process}
\label{models:sequential}

As mentioned previously, our sequential synthesis process synthesizes AvailableDays first and Price second. Moreover, AvailableDays is used as an additional predictor for the linear regression synthesis model for log Price (Equation (\ref{eq:7})). Such a sequential synthesis process requires fitting the synthesis models on the confidential data. Moreover, the synthesized values of variables are used in the synthesis steps of other variables where the synthesized variables serve as predictors (e.g., \citet{SynLBD2011}). In this section, we describe how we create $m > 1$ synthetic datasets and how our sequential synthesis works.

We use Just Another Gibbs Sampler (JAGS; \citet{JAGS}) for Markov Chain Monte Carlo (MCMC) estimation of the zero-inflated truncated Poisson regression model for AvailableDays and the linear regression model for Price. To create $m$ synthetic datasets, we generate $m$ sets of posterior parameter draws from $m$ independent MCMC iterations. For each set of posterior parameter draws, $\ell = 1, \cdots, m$, we first simulate the synthetic AvailableDays, denoted as $\tilde{Y}_{days,i}^{\ell}$, for the $i^{th}$ record in the $\ell^{th}$ synthetic dataset using Equation (\ref{eq:1}). Next we use the \emph{synthesized} $\tilde{Y}_{days,i}^{\ell}$ to generate the mean $\tilde{\mu}_i^{\ell}$ for log Price using,
\begin{equation} \label{eq:9}
    \tilde{\mu}_i^{\ell} = \gamma_{1}^{\ell}X_{1i} + \gamma_{2}^{\ell}X_{2i} + \gamma_{3}^{\ell}\cdot X_{1i} + \gamma_{4}^{\ell}\cdot \tilde{Y}_{days,i}^{\ell},
\end{equation}
where $\gamma_1^{\ell}, \cdots, \gamma_4^{\ell}$ refer to $\ell^{th}$ set of posterior parameter draws of these parameters. The resulting mean $\tilde{\mu}_i^{\ell}$ is then used to synthesize the synthetic Price variable $\tilde{Y}_{price,i}^{\ell}$ for the $i^{th}$ record in the $\ell^{th}$ synthetic dataset using Equation (\ref{eq:6}) and exponentiation. We repeat this process for every record $i$ in the New York Airbnb Open Dataset sample and the resulting $\ell^{th}$ synthetic dataset contains un-synthesized $(\bm X_1, \bm X_2, \bm X_3)$ as well as synthesized AvailableDays and Price $(\bm{\tilde{Y}}_{days}^{\ell}, \bm{\tilde{Y}}_{price}^{\ell})$.

The above process is then repeated for $m$ times to create $m$ synthetic datasets, \\ $\{(\bm X_1, \bm X_2, \bm X_3, \bm{\tilde{Y}}_{days}^{1}, \bm{\tilde{Y}}_{price}^{1}), \cdots, (\bm X_1, \bm X_2, \bm X_3, \bm{\tilde{Y}}_{days}^{m}, \bm{\tilde{Y}}_{price}^{m})\}$. In our case study, we choose $m = 20$ to adequately capture the uncertainty in the synthesis process as well as within reasonable computation time.

\section{Utility evaluation}
\label{Utility}

Data utility of synthetic data refers to how well the released synthetic data preserve the characteristics of the confidential data for analysis purposes. There are generally two types of utility measures: 1) global utility, which focuses on measuring the distance between the confidential data and the synthetic data; 2) analysis-specific utility, which emphasizes on how close the inferential results a user would obtain from the synthetic data are to those from the confidential data \citep{woo2009global, Snoke2018JRSSA}. In our investigation, we attempt both types of utility to provide a comprehensive evaluation of the utility performance of our synthesis.

Before presenting the detailed evaluation in Sections \ref{utility:global} and \ref{utility:specific}, we first visually inspect the synthetic data utility by plotting the synthetic AvailableDays and Price value compared to their corresponding confidential values. Figure \ref{fig:sub1} shows comparisons of histograms of the AvailableDays variable distribution for the confidential dataset and three synthetic datasets. In most bins in the histograms, the confidential AvailableDays has a similar count value with the synthetic values. However, in some bins, especially those towards the right tail of the distribution, the confidential AvailableDays deviates slightly from the synthetic values. Other synthetic datasets show similar results and are omitted for readability and brevity.

\begin{figure}[H]
\centering
\begin{subfigure}{.5\textwidth}
  \centering
  \includegraphics[width=1.0\linewidth]{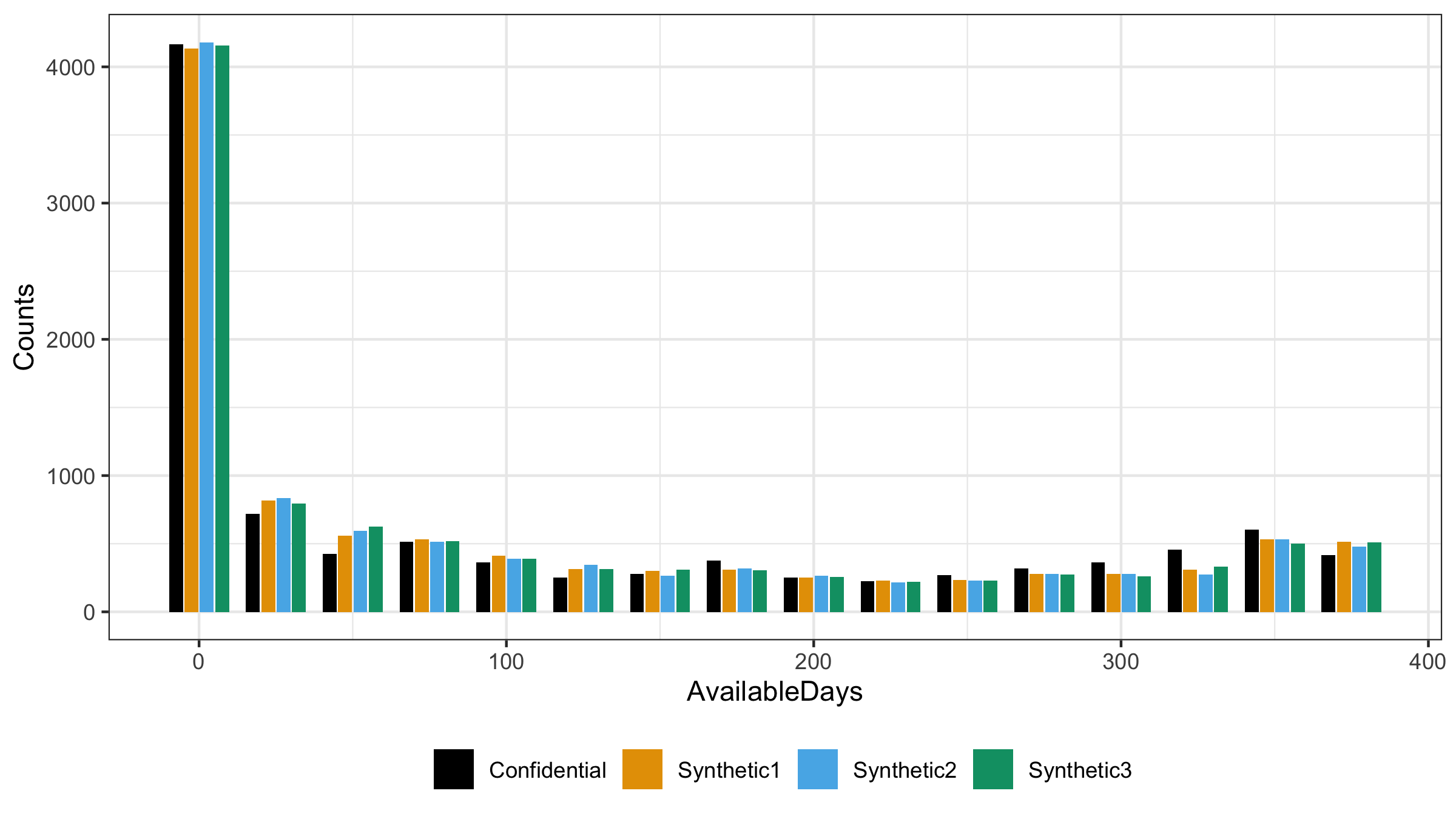}
  \caption{AvailableDays}
  \label{fig:sub1}
\end{subfigure}%
\begin{subfigure}{.5\textwidth}
  \centering
  \includegraphics[width=1.0\linewidth]{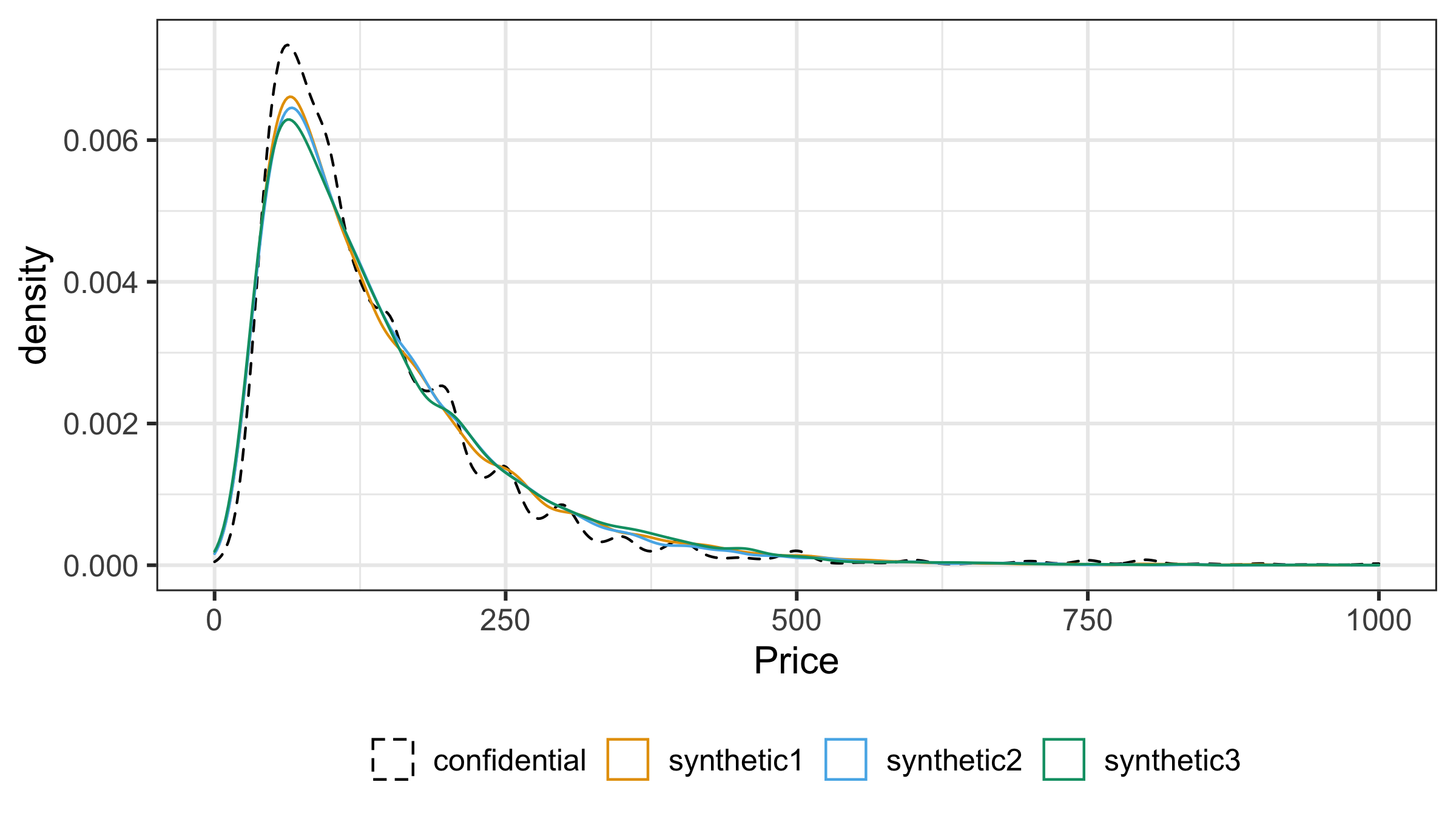}
  \caption{Price}
  \label{fig:sub2}
\end{subfigure}
\caption{Histograms comparing the confidential AvailableDays distribution and the synthetic AvailableDays distributions (left) and density plots comparing the confidential Price distribution and the synthetic Price distributions (right) from three synthetic datasets.The bin-width of the histograms is 25 days.}
\label{fig:AvailableDays_Price}
\end{figure}

As for Price, Figure \ref{fig:sub2} shows comparisons of density plots of the confidential Price distribution and the synthetic Price distributions from three synthetic datasets. The three synthetic Price distributions are highly similar to each other, all of which largely follow the confidential distribution of the confidential Price variable despite having lower peaks. It is also worth noting that some local features in the right tail of the confidential distribution are not adequately captured by the synthetic data distributions. Other synthetic datasets show similar results and are omitted for readability and brevity.

Overall, the synthetic AvailableDays and Price distributions largely preserve the characteristics of their corresponding confidential distributions. In general, we observe a decrease in distribution similarity for Price compared to AvailableDays. This can be partly attributed to the fact that we use the synthetic AvailableDays values to sequentially synthesize the Price variable. Therefore, we expect to observe a decrease in utility for synthetic Price values in our detailed evaluation presented in the sequel.

We now proceed to the global utility and analysis-specific utility evaluations. All results are based on evaluation of $m = 20$ generated synthetic datasets.

\subsection{Global utility}
\label{utility:global}

For global utility evaluation, we consider three different measures: 1) the propensity score measure; 2) the cluster analysis measure; and 3) the empirical CDF measure \citep{woo2009global, Snoke2018JRSSA}. All results are based on $m = 20$ synthetic datasets.

The resulting data utility measure evaluations for both the synthetic AvailableDays and Price variables are shown in Table \ref{tab:Data Utility}, where each column corresponds to a measure: 1) Propensity $U_p$ is the propensity score utility measure (multivariate); 2) Cluster $U_c$ is the cluster analysis utility measure (multivariate); 3) eCDF-max $U_m$ and eCDF-avg $U_a$ are the maximum absolute difference and squared average difference measures of the empirical CDF utility measure (univariate), respectively.

\begin{table}[t]
\centering
\begin{tabular}{>{\centering}p{1.1in} >{\centering}p{1in} >{\centering}p{0.95 in} >{\centering}p{1.2in} >{\centering}p{1.2in}}
\hline
  & Propensity $U_p$  & Cluster $U_c$ & eCDF-max $U_m$ & eCDF-avg $U_a$ \tabularnewline
\hline
AvailableDays & \multirow{2}{*}{0.00014} & \multirow{2}{*}{$4.63\cdot 10^{-5}$}  & 0.03140  & 0.00028  \tabularnewline
Price      & &  & 0.05090 & 0.00040 \tabularnewline \hline
High Utility & $\approx 0$  & $\approx 0$ & $\approx 0$  & $\approx 0$\tabularnewline
Low Utility & $\approx 1/4$  & large $U_c$ & large $U_m$ & large $U_a$\tabularnewline
\hline
\end{tabular}
\vspace{1mm}
\caption{Summary of the results of the propensity score, cluster analysis, and empirical CDF global utility measures.}
\label{tab:Data Utility}
\end{table}

The propensity score measure aims at distinguishing the confidential and synthetic datasets using a classification algorithm. The reported propensity score is the average of the squared distance from the estimated propensity score for each record from a constant $c$, the percentage of synthetic values in the merged dataset of the confidential and synthetic values \citep{woo2009global, Snoke2018JRSSA}. In our evaluation, we calculate the propensity score measure on each synthetic dataset using a logistic regression as the classification algorithm, where the main effects of all variables are used as predictors. Since each synthetic dataset has the same size as the confidential dataset (a feature of partial synthesis; therefore $c = 1/2$), a propensity score measure close to 0 suggests high degree of similarity between the synthetic and confidential data, which implies high data utility, whereas a propensity score measure close to $1/4$ indicates low data utility. We take the average measure across the $m = 20$ synthetic datasets and the resulting value is $U_p = 0.00014$. This result suggests that the synthetic datasets are similar to the confidential dataset and thus the synthetic dataset has high data utility.

The cluster analysis measure groups the merged dataset of the confidential dataset and a synthetic dataset into multiple clusters using a clustering algorithm \citep{woo2009global}. This measure evaluates data utility based on comparing the number of confidential records and synthetic records in each cluster. The reported $U_c$ is the weighted sum of differences in proportions of synthetic records in each cluster and those of synthetic records in the entire merged dataset (which is $1/2$ in our case study). Therefore, a $U_c$ measure close to $0$ indicates the clustering algorithm cannot distinguish the synthetic data from the confidential data in each cluster, suggesting higher data utility. In our evaluation, we use the the unweighted pair group method with arithmetic mean (UPGMA, \citet{sokal1958statistical}) method as our clustering algorithm and report the average $U_c$ value across $m = 20$ synthetic datasets. The reported $U_c = 4.63\cdot 10^{-5}$, as shown in Table \ref{tab:Data Utility}, is very small and close to $0$. Therefore, the cluster analysis utility measure indicates high data utility of our simulated synthetic datasets.

Last but not least, the empirical CDF measure compares the empirical cumulative distribution functions (CDF) of the marginal distribution of one variable between the confidential and the synthetic data. The utility of one synthetic dataset is evaluated by whether the empirical CDF measure can discriminate the confidential dataset from this synthetic dataset \citep{woo2009global}. The empirical CDF utility has two components: 1) $U_m$, the maximum absolute difference between the empirical CDFs of the confidential dataset and the synthetic dataset; 2) $U_a$, the average squared difference between the two empirical CDFs. Both $U_m$ and $U_a$ are non-negative. Small $U_m$ and $U_a$ values indicate high similarity between the the two empirical CDFs, suggesting high synthetic data utility. Conversely, large $U_m$ and $U_a$ values indicate low data utility. We report the average $U_m$ and $U_a$ values across $m = 20$ synthetic datasets. As shown in Table \ref{tab:Data Utility}, the $U_m$ measure values for the AvailableDays and Price variables are $U_m = 0.03140$ and $U_m=0.05090$ respectively. The $U_a$ measure values for these two variables are $U_a = 0.00028$ and $U_a = 0.00040$ respectively. Both the maximum absolute difference and the average difference between the empirical CDFs of the confidential and synthetic datasets are close to $0$, suggesting that the confidential and the synthetic data have similar CDFs, indicating high data utility.

In summary, our global utility measures of propensity score, cluster analysis, and empirical CDFs all suggest a high level of utility preservation of our synthetic  datasets for the New York Airbnb Open Data sample. We next present analysis-specific utility measures on several quantities of interest.

\subsection{Analysis-specific utility}
\label{utility:specific}

To further evaluate the data utility of the synthetic datasets, we perform analysis-specific utility measures to compare a number of quantities of interest, including the mean estimate, several extreme quantile estimates, and some regression coefficient estimates. Since our synthetic datasets are partially synthetic, we obtain the point estimate and 95\% confidence interval estimates of the mean and regression coefficients following the partial synthesis combining rules \citep{reiter2003SM, Drechsler2011book}. For quantile estimates, we use bootstrapping methods \citep{HuSavitskyWilliams2021JSSAM}.

After calculating the 95\% confidence interval estimates from the confidential data and the synthetic data, we use the interval overlap measure to quantify their distance \citep{karr2006framework, Snoke2018JRSSA}. Specifically, for the $\ell^{th}$ synthetic dataset, we calculate the interval overlap measure, $I^{\ell}$, for a quantity of interest, such as the mean, as the following:
\begin{equation}\label{eq:10}
    I^{\ell} = \frac{U_i^{\ell} - L_i^{\ell}}{2(U_c - L_c)} + \frac{U_i^{\ell} - L_i^{\ell}}{2(U_s^{\ell} - L_s^{\ell})},
\end{equation}
where $U_c$ and $L_c$ are the upper and lower bound of the $95\%$ confidence interval obtained from the confidential dataset, $U_s^{\ell}$ and $L_s^{\ell}$ are the upper and lower bound of the $95\%$ confidence interval obtained from the $\ell^{th}$ synthetic dataset. Moreover, $U_i^{\ell} = \min(U_c, U_s^{\ell})$, and $L_i^{\ell} = \max(L_c, L_s^{\ell})$, where subscript $i$ refers to intersection (of the two confidence intervals). Finally, we report the average $I = 1 / m \sum_{\ell = 1}^{m} I^{\ell}$ as the interval overlap for the quantity of interest. An $I$ value close to $1$ indicates that a high degree of overlap, suggesting high data utility of the synthetic datasets. In contrast, an $I$ value close to $0$ or even negative indicates little or no overlap, suggesting poor data utility of the synthetic datasets.

We first consider the point estimate, the 95\% confidence interval, and the interval overlap measures for AvailableDays and Price, where we report those for the mean, the 25\% quantile, and the 90\% quantile. Results for the 90\% quantile are included in Figure \ref{fig:Q90_CI} whereas the other ones are included in the Supplementary Materials for brevity. In Figure \ref{fig:Q90_CI}, panel A is for AvailableDays and panel B is for Price. In each panel, we include the point estimate (box with a cross) and the 95\% confidence interval (the vertical bars) of the confidential data (black and left) and of the synthetic data (yellow and right). Moreover, we add the interval overlap measure $I$ for each quantity in the panel.

\begin{figure}[htb!]
    \centering
    \includegraphics[scale = 0.095]{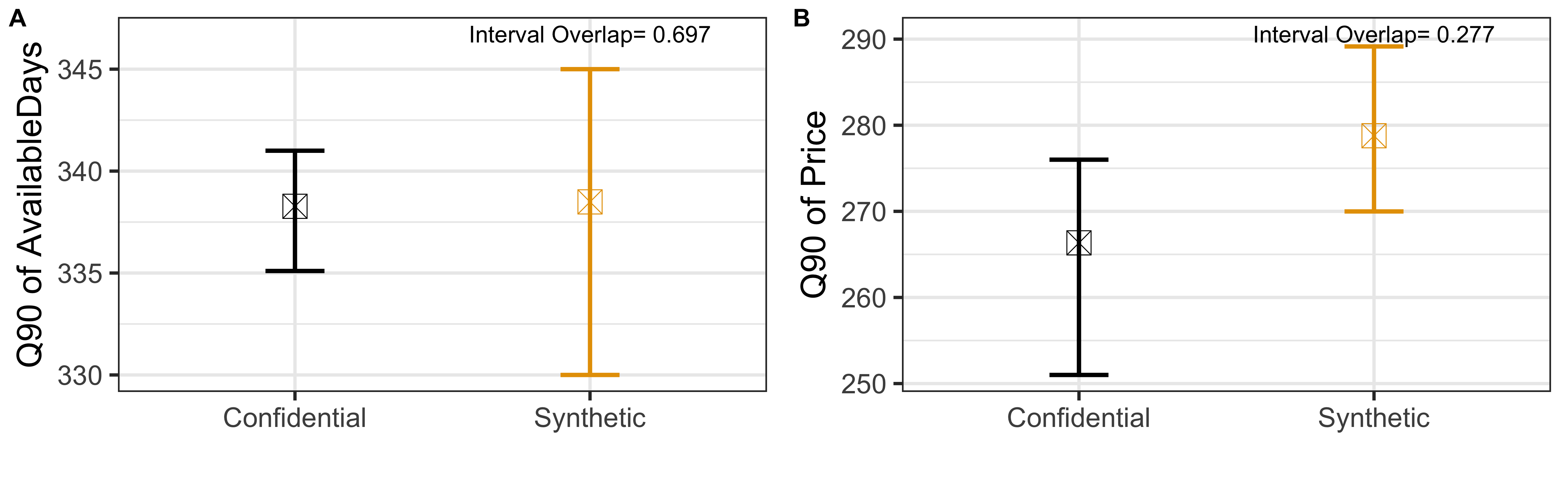}
    \caption{The point estimate, $95\%$ confidence interval, and interval overlap of the $90\%$ quantiles of the confidential and synthetic datasets.  Panel \textbf{A} shows the estimand for AvailableDays and panel \textbf{B} shows the estimand for Price.}
    \label{fig:Q90_CI}
\end{figure}

Overall, with the exception for the mean of AvailableDays, the point estimates of the three quantities of interest are close between the confidential data and the synthetic data for AvailableDays, and the 95\% confidence interval overlap measures are close to 1 as well, suggesting a high level of utility preservation. For the 25\% and 90\% quantile estimates, Price are slightly worse, which are expected since Price is synthesized after AvailableDays is synthesized according to the sequential synthesis process. These results suggest that when using the synthetic data, users would obtain point estimates and 95\% confidence interval estimates for key quantities of AvailableDays and Price that are overall close to those from the confidential data.

\begin{figure}[htb!]
    \centering
    \includegraphics[scale = 0.095]{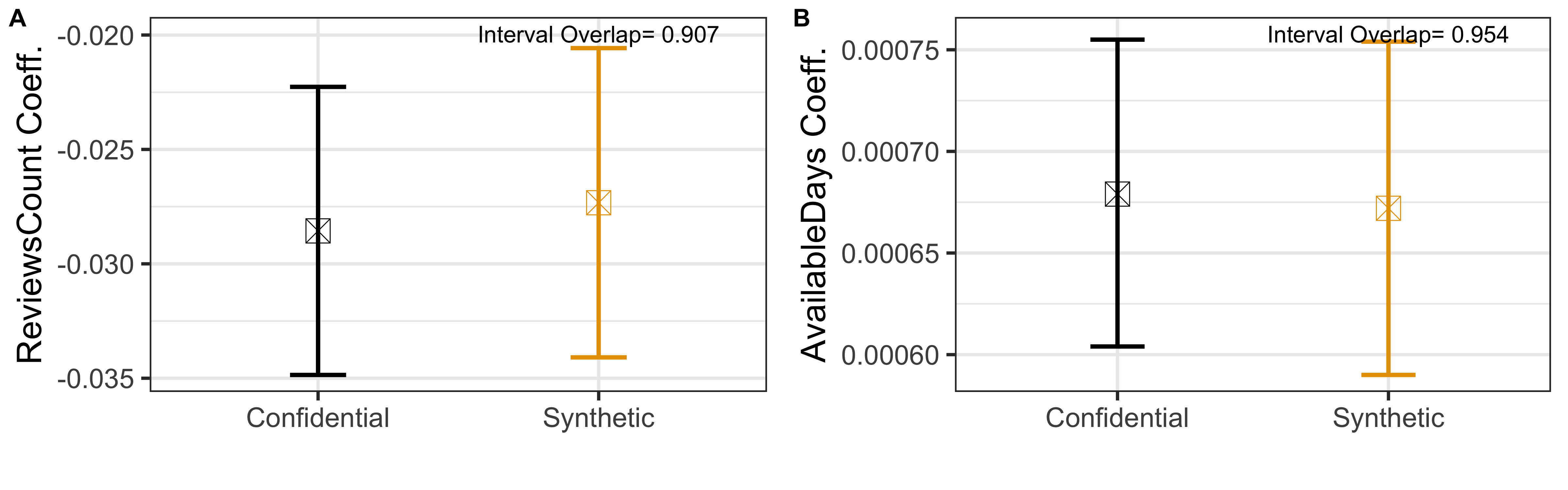}
    \caption{The point estimate, $95\%$ confidence interval, and interval overlap of selected regression coefficients from a linear regression analysis from the confidential and synthetic datasets. Panel \textbf{A} shows the ReviewsCount coefficient and panel \textbf{B} shows the AvailableDays coefficient.}
    \label{fig:regression_CI}
\end{figure}

We also consider regression analysis types of analysis-specific utility measures. Specifically, we examine the regression coefficients of the effects of Neighborhood, RoomType, ReviewsCount, and AvailableDays on the response variable, Price, using a linear regression. Recall that both AvailableDays and Price are synthesized, therefore this regression analysis utility measure takes into account all synthesized variables. As with the mean estimates, partial synthesis combining rules are applied in obtaining the regression coefficients from the $m = 20$ synthetic datasets. For each coefficient, we have a point estimate and a 95\% confidence interval estimate. Given the confidence interval estimates from the confidential data and the synthetic data, we calculate the interval overlap measure. Selected results are displayed in Figure \ref{fig:regression_CI}.

The point estimates of the coefficients for ReviewsCount and AvailableDays given the synthetic data are close to those given the confidential data. In both cases, the interval overlap measures are almost 1, indicating an extremely high level of utility. Results for the other regression coefficients show similarly high utility with interval overlap measures above $0.9$ for most of them. These are omitted for brevity. 

In summary, our analysis-specific utility measures on key quantities of interest, such as means, quantiles, and regression coefficients, all suggest a high level of utility preservation of the generated synthetic datasets for the New York Airbnb Open Data sample. Combined with the results of global utility evaluation in Section \ref{utility:global}, we conclude that the synthetic datasets generated from our proposed sequential synthesis models are able to preserve important characteristics of the confidential New York Airbnb Open Data sample. We now proceed to evaluate the disclosure risks associated with these synthetic data. 

\section{Disclosure risk evaluation}
\label{Risk}

The utility evaluation results presented in Section \ref{Utility} demonstrate a high level of utility preservation of our synthetic datasets. However, synthetic data should only be considered for public release if their level of privacy protection is satisfactory. In this section, we assess the disclosure risks associated with our simulated synthetic datasets. Since our synthesis is partial synthesis of AvailableDays and Price, we consider attribute disclosure risk and identification disclosure risk, both of which exist in partially synthetic data \citep{hu2019bayesian}.

\subsection{Attribute disclosure risk}

Attribute disclosure risk refers to the risk that a data intruder correctly infers the confidential values of sensitive, synthesized variables given the released synthetic dataset and other information from external databases. In our case study, AvailableDays and Price are synthesized whereas RoomType, Neighborhood, and ReviewsCount are kept un-synthesized. If an intruder has access to information on records' information on some or all of these three un-synthesized variables, attribute disclosure could occur.

In our evaluation, we measure the attribute disclosure risk by estimating the sum of the average probabilities that the attributes of the sensitive variables are exposed for each record in the confidential dataset. Our approach is similar to the one proposed by \citet{mitra2020confidentiality}. The process is applied to each of the $m$ synthetic datasets and we drop the synthetic dataset index $\ell$ for notation simplicity in this section. Once the probabilities are calculated for synthetic dataset $\ell$, we calculate and report the average across $m$ synthetic datasets as the attribute disclosure risk measure. Results are based on $m = 20$ synthetic datasets.

For record $i$, we first identify all records in the synthetic dataset that are matched with record $i$ based on the un-synthesized variables. Denote this set of matched records as $\mathcal{M}_i$. Next, we calculate the probability that the attributes of record $i$ are exposed to the intruder, which is defined as the fraction of records that share similar sensitive attributes among the matched set of records in $\mathcal{M}_i$. In our case study, we use the un-synthesized RoomType, Neighborhood, and ReviewsCount for the matching step and consider similar sensitive attributes for synthesized AvailableDays and Price. Specifically, we consider records in $\mathcal{M}_i$ that have AvailableDays within $\pm 5$ or $\pm 10$ days and Price within $\pm 5\%$ or $\pm 10\%$ of Price of record $i$ as similar, following the choices made in \citet{mitra2020confidentiality}.

Mathematically, the attribute disclosure risk probability of record $i$ can be expressed as the following:
$$
p_i = \frac{1}{c_i}\sum_{j\in \mathcal{M}_i}I\left( |\text{AvailableDays}_j - \text{AvailableDays}_i| \le \text{r\_avail}, \frac{|\text{Price}_j - \text{Price}_i|}{\text{Price}_i} \le \text{r\_price}  \right),
$$
where $I()$ is the indicator function that takes value 1 if the condition is met and 0 otherwise, $c_i$ is the number of records in $\mathcal{M}_i$, r\_avail and r\_price are the radius within which records are considered similar, and finally $p_i$ represents the attribute disclosure risk probability for record $i$. In our case study, we consider the absolute difference between AvailableDays with a radius of $\text{r\_avail} \in \{5, 10\}$ and the percentage absolute difference between Price with a radius of $\text{r\_price} \in \{5\%, 10\%\}$.

We denote the final attribute disclosure risk measure of the entire synthetic dataset by $AR = \sum_{i=1}^{n} p_i$, and we compare and report the attribute disclosure risk of the synthetic and confidential datasets under different r\_avail and r\_price choices, shown in Table \ref{tab:AR Table}. 

\begin{table}[h]
\centering
\begin{tabular}{>{\centering}p{0.7in} >{\centering}p{0.7in} >{\centering}p{2 in} >{\centering}p{1.7 in}}
\hline
r\_avail & r\_price & Confidential AR & Synthetic AR \tabularnewline
\hline
5 & 5\% & 636.40 & 126.07 \tabularnewline
\hline
10 & 5\% & 657.53 & 146.44 \tabularnewline
\hline
10 & 10\% & 816.55 & 291.73  \tabularnewline
\hline
\end{tabular}
\vspace{1mm}
\caption{The attribute disclosure risk (AR) of the confidential dataset and the synthetic datasets under several (r\_avail, r\_price) combinations.}
\label{tab:AR Table}
\end{table}

As evident in the last column of Table \ref{tab:AR Table}, the attribute disclosure risk of the synthetic datasets increases as the radius choices increase. This result is intuitive since a more stringent attribute similarity definition (i.e., smaller radius values) will result in a lower probability of finding similar attributes in the synthetic datasets. We also calculate the inherent attribute disclosure risk of the confidential dataset, shown in the third column of Table \ref{tab:AR Table}. The comparison of these two columns shown that the data synthesis process has achieved a significant reduction (3 to 4 folds) in the attribute disclosure risk.

\subsection{Identification disclosure risk}
Identification disclosure occurs when a data intruder, with certain knowledge, correctly establishes a one-to-one relationship between a record in the confidential data and this record in the synthetic data. Given the nature of our synthesized variables, count for AvailableDays and continuous for Price, we follow the approaches in \citet{hornby2021TDP} when declaring a match between a record in the confidential data with another record in the synthetic data. Specifically, among all records sharing the same un-synthesized RoomType, Neighborhood, and ReviewsCount in the confidential and synthetic data, a match is declared for confidential record $i$ if for a synthetic record $j$, its synthetic AvailableDays value is within $5$ days from the confidential AvailableDays value of record $i$ and if the log transformation of its synthetic Price value is within the $\pm 5\%$ range of the confidential $\log{\textrm{(Price)}}$ value of record $i$. The choices of $\pm 5$ days for AvailableDays and $\pm 5\%$ for Price are based on previous studies \citep{mitra2020confidentiality} and the Airbnb data sample. 
A thorough investigation of the choices would be an interesting future research direction. As before, evaluation is performed on each synthetic dataset separately and results are combined by taking averages. All results are based on $m = 20$ synthetic datasets.

The identification disclosure risks of the confidential and the synthetic data are evaluated based on three commonly-used measures, namely the expected match risk (EMR), true match rate (TMR), and false match rate (FMR) \citep{reiter2009estimating, hu2019bayesian, DrechslerHu2021JSSAM}, presented in Table \ref{tab: IR_table}. 
The EMR represents the expected number of records in the synthetic datasets to be correctly identified. The TMR refers to the percentage of true unique matches among all records, and the FMR measures the percentage of false unique matches respectively among all unique matches. Therefore, lower EMR, lower TMR, and higher FMR indicate lower identification risks, and vice versa. 




\begin{table}[h]
\centering
\begin{tabular}{>{\centering}p{1.2in} >{\centering}p{1in} >{\centering}p{1in} >{\centering}p{1in} >{\centering}p{1in}}
\hline
 & EMR & TMR & FMR & u \tabularnewline
\hline
Confidential IR & 7182.03 & 0.62 & 0.00 & 6226 \tabularnewline
\hline
Synthetic IR & 125.59 & 0.01 & 0.91 & 1153 \tabularnewline
\hline
\end{tabular}
\vspace{1mm}
\caption{Identification disclosure risk (IR) of the confidential data and synthetic data, where EMR stands for the expected match risk, TMR stands for the true match rate, FMR stands for the false match rate, and u is the number of unique matches.}
\label{tab: IR_table}
\end{table}


Overall results in Table \ref{tab: IR_table} suggest low identification risks associated with our synthetic datasets for the New York Airbnb Open Data sample. Indeed, compared to the inherent identification disclosure risk of the confidential data, the synthetic data show a striking reduction in the probability that a record will be correctly identified by data intruder by more than 50 folds (i.e., 0.7182 vs 0.0126, as calculated by the values in the EMR column in Table \ref{tab: IR_table} divided by $n = 10,000$ in the sample). This evaluation is based on the assumption that the intruder would have perfect knowledge of the un-synthesized variables, RoomType, Neighborhood, and ReviewsCount, which could be unrealistic in practice and potentially overstate the disclosure risk concerns. To evaluate the impact of uncertainties in intruder's knowledge, we proceed to incorporate scenarios of such uncertainties in the next section.

\subsection{Uncertainties in intruder's knowledge in identification disclosure risk evaluation}

To investigate how uncertainties in intruder's knowledge would influence the identification disclosure risks of the synthetic data, we consider adding normal noise to the ReviewsCount variable after applying a log transformation. This is to mimic uncertainties in intruder's knowledge about this ReviewsCount variable. Specifically, for record $i$,
\begin{equation}
   \log(\textrm{ReviewsCount}^*_{i}) \sim \textrm{Normal}(\log(\textrm{ReviewsCount}_{i}), S \times \textrm{ReviewsCount}_{i}), 
\end{equation}
where $\textrm{ReviewsCount}^*_{i}$ represents the intruder's knowledge of the ReviewsCount value, $\textrm{ReviewsCount}_{i}$ is the confidential ReviewsCount value, and $S \times \textrm{ReviewsCount}_{i}$ represents the level of uncertainties in intruder's knowledge, which is the standard deviation of the normally added noise. We experiment with the percentage $S$ from $\{0.00, 0.01, \cdots, 0.15\}$ and evaluate the identification disclosure risks under each different $S$ value as the uncertainty level of ReviewsCount. As before, results of EMR, TMR, and FMR are reported with each $S$ value. We also report the number of unique matches in each case.

\begin{figure}[htb!]
    \centering
    \includegraphics[scale = 0.14]{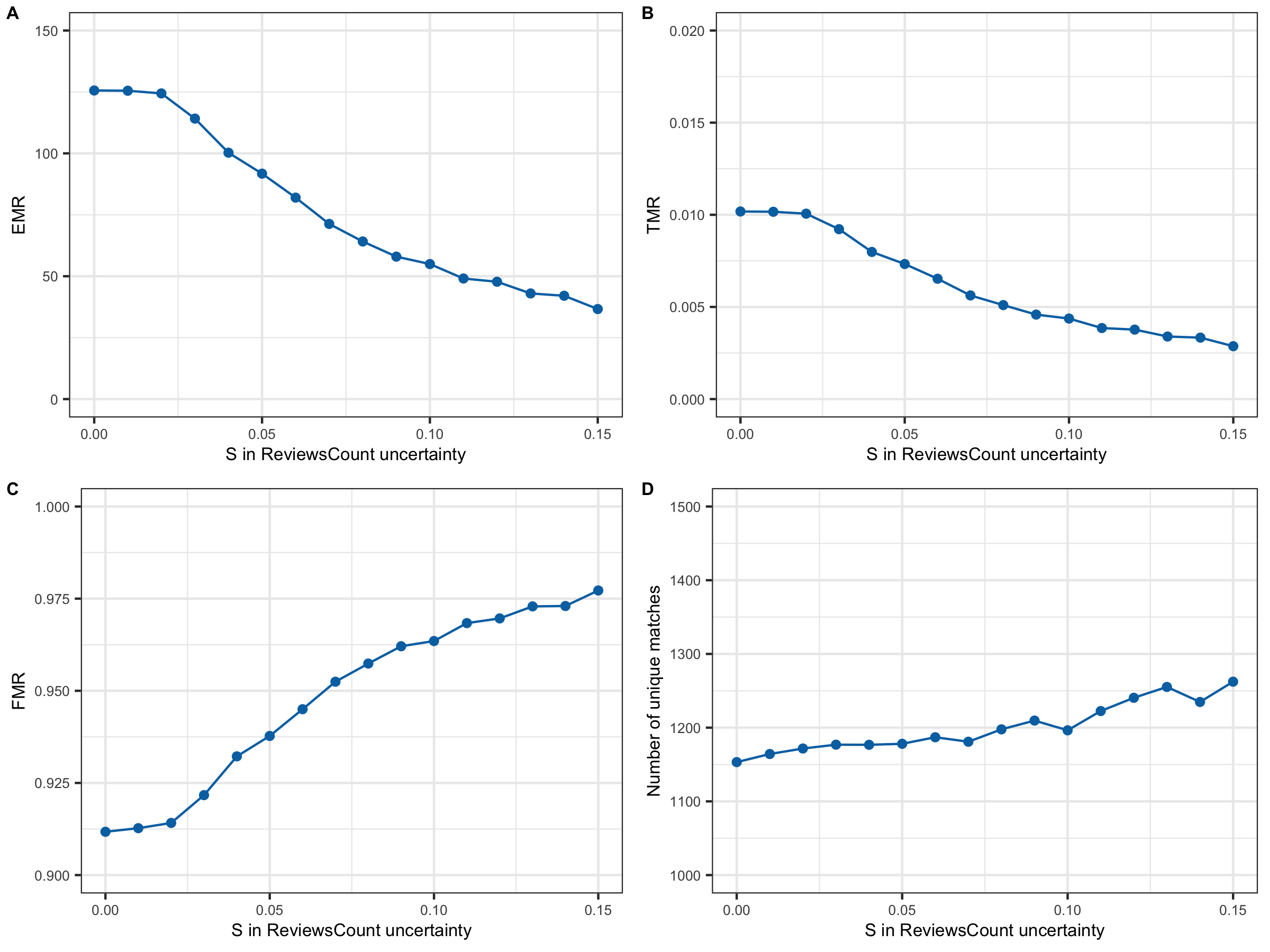}
    \caption{Identification disclosure risks for the synthetic datasets under different standard deviation percentage value ($S$) of uncertainties in intruder knowledge of ReviewsCount. Panel \textbf{A} shows the impact of $S$ on EMR, panel \textbf{B} displays that on TMR, and panel \textbf{C} presents that on FMR. Panel \textbf{D} plots impacts of $S$ on the number of unique matches.}
    \label{fig:IR}
\end{figure}

Results are presented in Figure \ref{fig:IR} for all different experimented $S$ values. Evidently, as the level of uncertainties increases (i.e., $S$ increases), both EMR and TMR decrease, while FMR increases. Therefore, in general, as data intruder's knowledge becomes more uncertain, the identification disclosure risks for the synthetic datasets decrease, a not-so-surprising result. These can be explained by the fact that the uncertainties in ReviewsCount make it more difficult for the intruder to establish a one-to-one relationship between two records, one in the confidential data and the other in the synthetic data. However, for the same reason, uncertainties in intruder's knowledge would reduce the inherent identification disclosure risk of the confidential dataset. Nevertheless, for all uncertainty levels $S$ we considered, the synthetic data offer significant improvement in privacy protection by reducing the probability of identifying a record by at least 50 folds. Results are included in the Supplementary Materials for further reading.

It is worth noting that while panel \textbf{D} in Figure \ref{fig:IR} shows that in general the number of unique matches in the synthetic datasets increases as $S$ increases, the FMR rises rapidly with as the amount of uncertainties in intruder knowledge increase. These results suggest that uncertainties in ReviewsCount increase the probability of a unique match to be a false match, and higher uncertainty level makes it harder for data intruder to correctly identify records in the confidential data with the released synthetic data.

\section{Conclusion}\label{Conclusion}

As synthetic data become more widely used by data disseminators, our case study focuses on how to best disseminate a sample of the New York Airbnb Open Data, where the number of available days and the price of Airbnb listings are considered sensitive information to be protected. Given the features of the number of available days, we propose a zero-inflated truncated Poisson regression model. Moreover, we utilize a sequential synthesis approach to further synthesize the price variable after synthesizing the available days. The resulting multiple partially synthetic datasets are evaluated for their utility and privacy protection performances.

Multiple utility measures, including global utility and analysis-specific utility, are evaluated for our synthetic New York Airbnb Open Data samples. Overall, our synthetic datasets preserve a high level of utility that can be hugely beneficial for users when conducting analyses. Attribute disclosure risk and identification disclosure risk are evaluated for our synthetic datasets under certain intruder's knowledge assumptions. Overall, there exist low disclosure risks in our synthetic datasets, making them suitable for public release.

An important methodological contribution of our work lies in our investigation of uncertainties in intruder's knowledge when evaluating identification disclosure risks. Specifically, we consider how the uncertainty level of intruder's knowledge of an un-synthesized variable impacts several summaries of identification disclosure risks. Our results show that as intruder's knowledge becomes more inaccurate (i.e., more uncertain), it is less likely for the intruder to correctly identify records in the confidential data with access to the released synthetic data and external information. 

Extending the uncertainty evaluation to attribute disclosure risks is an important and highly relevant future work direction. As with identification disclosure risk evaluation, assumptions are made when evaluating attribute disclosure risks. Investigating the impact of intruder's knowledge uncertainty in attribute disclosure risks would shed light on how to most realistically evaluate disclosure risks in synthetic data.

\bigskip
\section*{Acknowledgement}
The authors thank Joerg Drechsler for his expertise on the CART synthesis method.

\bibliography{ref.bib}

\section*{Supp 1: Additional Information about the New York Airbnb Open Dataset from Section 1.1}

\begin{table}[H]
\centering
\begin{tabular}{>{\centering}p{1in} >{\centering}p{1in} >{\centering}p{1 in} >{\centering}p{1in} >{\centering}p{1in}}
\hline
  & Mean & 25\% Quantile & Median & 75\% Quantile   \tabularnewline
\hline
AvailableDays & 113.044 & 0 & 43  & 228 \tabularnewline
Price      & 152.036 & 69 & 105 & 177 \tabularnewline
\hline
\end{tabular}
\vspace{1mm}
\caption{Summary statistics of the confidential AvailableDays and Price values.}
\label{tab:confidential data summary}
\end{table}

\section*{Supp 2: Additional Utility Results from Section 3.2}

\begin{figure}[H]
    \centering
    \includegraphics[scale = 0.095]{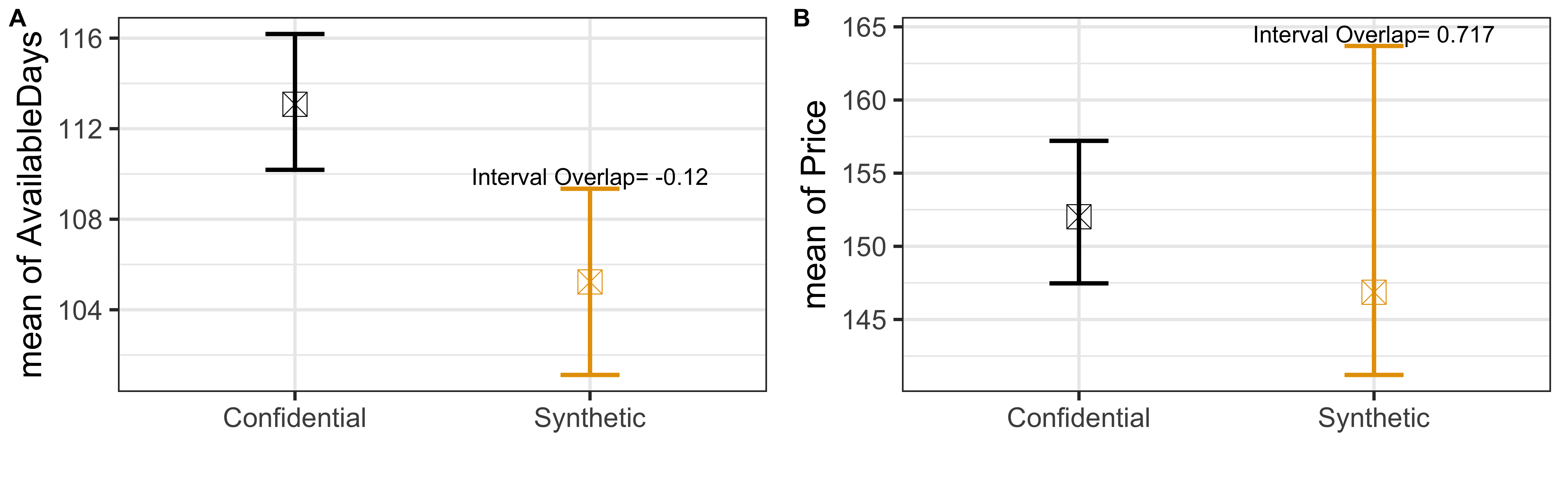}
    \caption{The point estimate, $95\%$ confidence interval, and interval overlap of the means of the confidential and synthetic datasets. Panel \textbf{A} shows the estimand for AvailableDays and panel \textbf{B} shows the estimand for Price.}
    \label{fig:mean_CI}
\end{figure}

\begin{figure}[H]
    \centering
    \includegraphics[scale = 0.095]{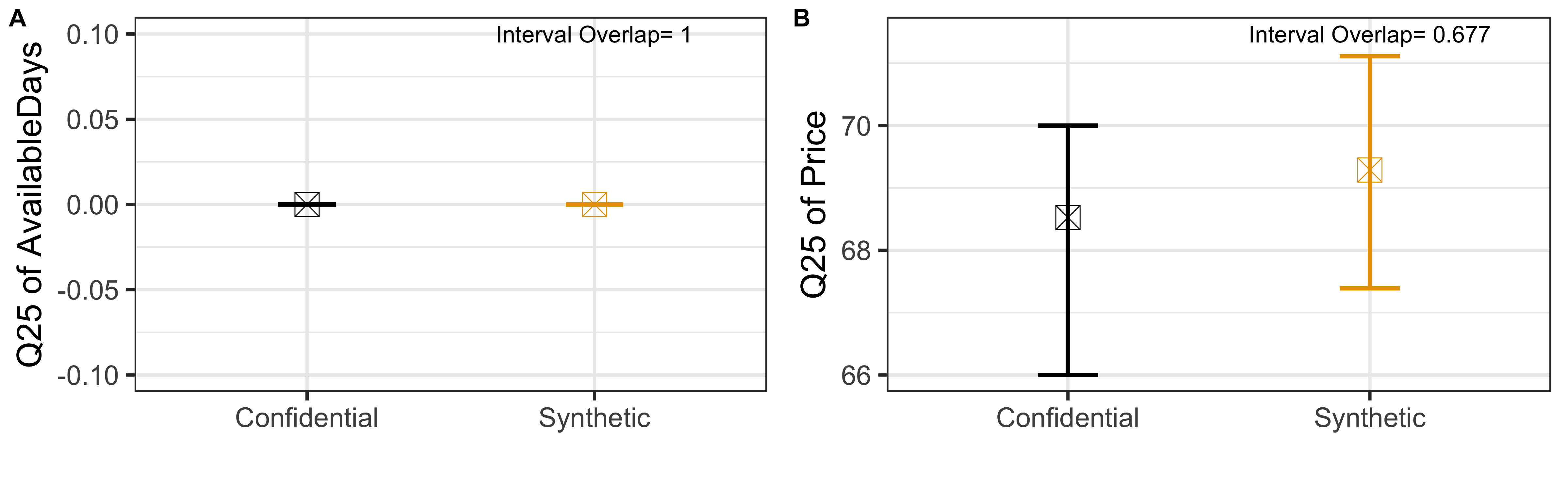}
    \caption{The point estimate, $95\%$ confidence interval, and interval overlap of the $25\%$ quantiles of the confidential and synthetic datasets.  Panel \textbf{A} shows the estimand for AvailableDays and panel \textbf{B} shows the estimand for Price.}
    \label{fig:Q25_CI}
\end{figure}

\section*{Supp 3: Additional Risk Results from Section 4.3}

\begin{table}[h]
\centering
\begin{tabular}{>{\centering}p{1.1in} >{\centering}p{0.9in} >{\centering}p{0.8in} >{\centering}p{0.8in} >{\centering}p{0.8in} >{\centering}p{0.8in}}
\hline
Uncertainty & Data & EMR & TMR & FMR & u \tabularnewline
\hline
\multirow{2}{*}{$1\%$} & Confidential & 7178.26 & 0.622 & 0 & 6220 \tabularnewline
& Synthetic & 125.49 & 0.010 & 0.913 & 1164 \tabularnewline
\hline
\multirow{2}{*}{$5\%$} & Confidential & 5061.25 & 0.438 & 0.073 & 4724 \tabularnewline
& Synthetic & 91.75 & 0.007 & 0.938 & 1178 \tabularnewline
\hline
\multirow{2}{*}{$10\%$} & Confidential & 2983.67 & 0.259 & 0.216 & 3298 \tabularnewline
& Synthetic & 54.96 & 0.004 & 0.964 & 1196 \tabularnewline
\hline
\multirow{2}{*}{$15\%$} & Confidential & 2094.65 & 0.182 & 0.325 & 2703 \tabularnewline
& Synthetic & 36.65 & 0.003 & 0.977 & 1262 \tabularnewline
\hline
\end{tabular}
\vspace{1mm}
\caption{Comparison of identification disclosure risks between the confidential and the synthetic datasets under selected intruder knowledge uncertainty levels.}
\label{tab:IR Uncertainty Compare}
\end{table}

\section*{Supp 4: The CART Synthesis Method and Result Comparisons}
This section compares the all-purpose synthesis method classification and regression trees (CART) with our proposed Bayesian synthesis models. We compare the two synthesis methods from the perspective of their respective data utility and disclosure risks of the resulting synthetic data.

\subsection*{Supp 4.1: Utility Results and Comparisons}
To compare the synthetic data generated by the CART and the Bayesian model, we first visually inspect the synthetic data utility by plotting the synthetic AvailableDays and Price value compared to their corresponding confidential values for the CART and the Bayesian model respectively.

\begin{figure}[H]
\centering
\begin{subfigure}{.5\textwidth}
  \centering
  \includegraphics[width=1.0\linewidth]{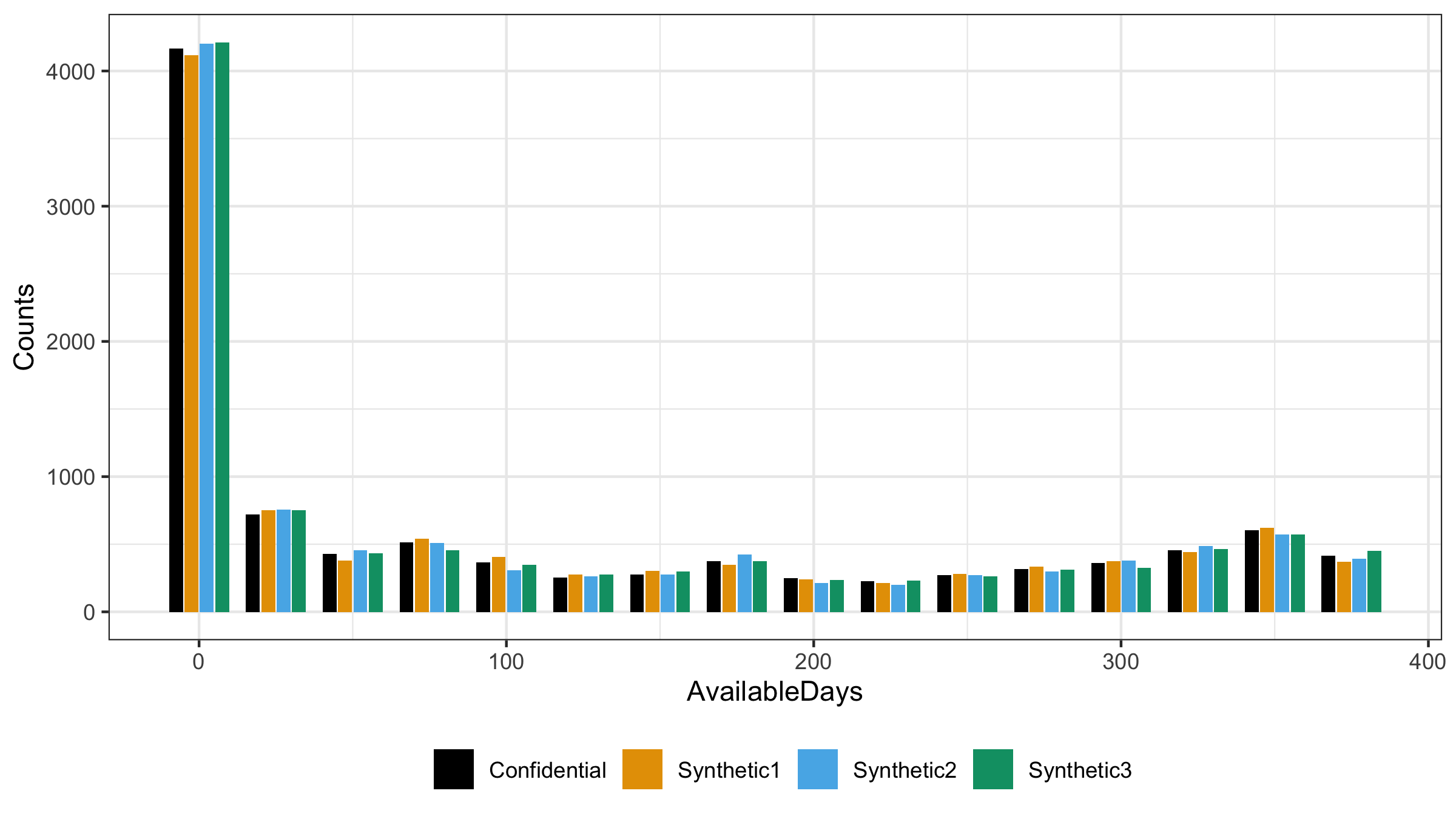}
  \caption{CART}
  \label{fig:sub1}
\end{subfigure}%
\begin{subfigure}{.5\textwidth}
  \centering
  \includegraphics[width=1.0\linewidth]{Original_vs_Synthetic_AvailableDays_histogram_3syn_bottom_legend.png}
  \caption{Bayesian}
  \label{fig:sub2}
\end{subfigure}
\caption{Histograms comparing the confidential AvailableDays distribution and the synthetic AvailableDays distributions from three CART-synthesized datasets and three Bayesian synthesized datasets. The bin-width is 25 days.}
\label{fig:AvailableDays_CART_Compare}
\end{figure}

Figure \ref{fig:AvailableDays_CART_Compare} compares the distributions of the synthesized AvailableDays variable from the CART method and from the Bayesian model. Figure \ref{fig:sub1} shows how three of the CART synthesized datasets' AvailableDays values compare to the confidential value and Figure \ref{fig:sub2} shows how three of the Bayesian synthesized datasets' AvailableDays values compare to the confidential value.

For both synthesis methods, the synthetic AvailableDays have similar count values as the confidential AvailableDays variable. For the Bayesian model, in some bins, especially those towards the right tail of the distribution, the synthetic AvailableDays deviates slightly from the confidential values. In comparison, CART generated AvailableDays show slightly better similarity with the confidential values.

\begin{figure}[H]
\centering
\begin{subfigure}{.5\textwidth}
  \centering
  \includegraphics[width=1.0\linewidth]{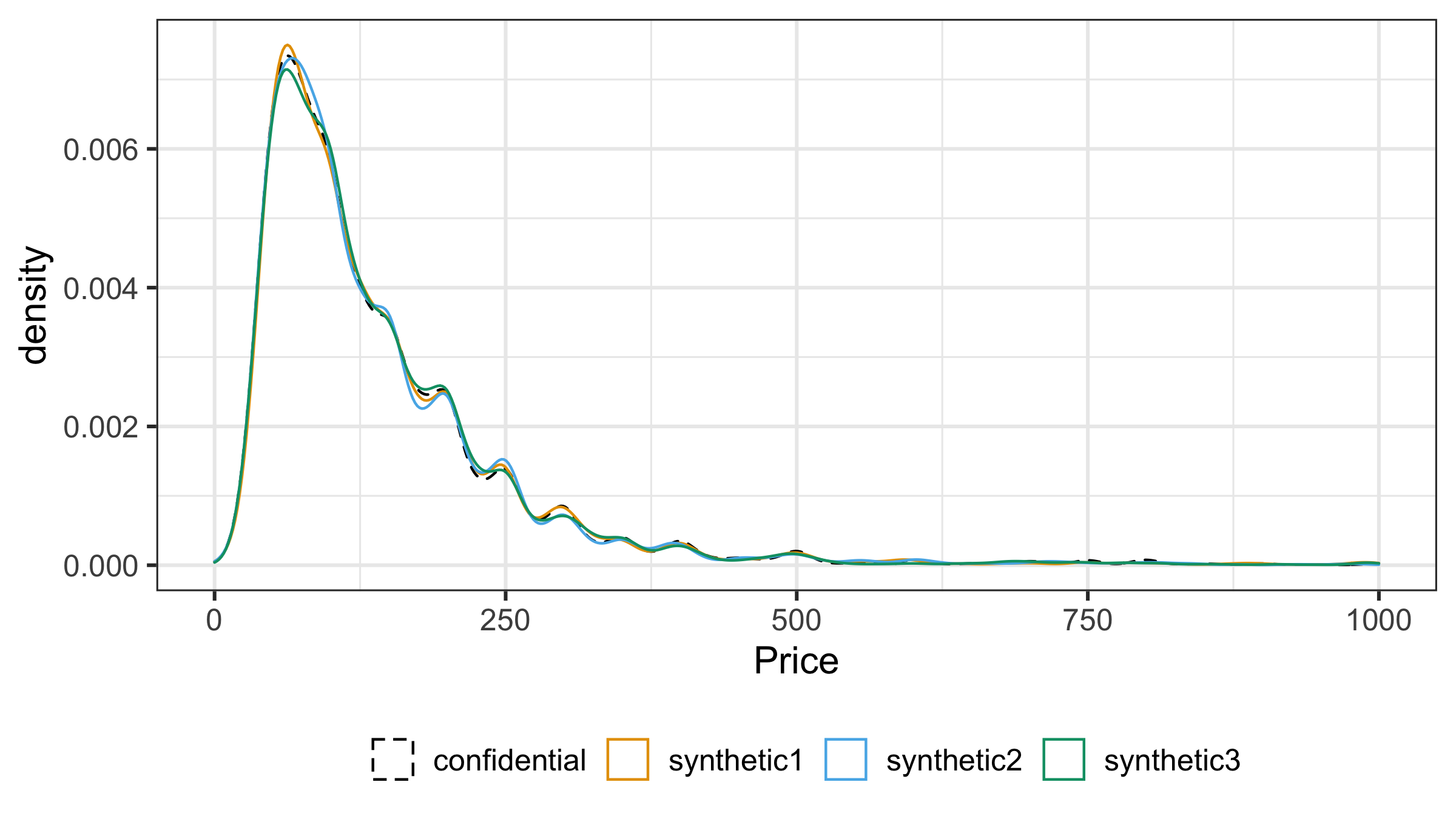}
  \caption{CART}
  \label{fig:sub3}
\end{subfigure}%
\begin{subfigure}{.5\textwidth}
  \centering
  \includegraphics[width=1.0\linewidth]{Original_vs_Synthetic_Price_Line_3syn.png}
  \caption{Bayesian}
  \label{fig:sub4}
\end{subfigure}
\caption{Density plot comparing the confidential Price distribution and the synthetic Price distributions from three CART-synthesized datasets and three Bayesian synthesized datasets.}
\label{fig:Price_CART_Compare}
\end{figure}

Figure \ref{fig:Price_CART_Compare} shows the comparisons of density plots of the confidential Price distribution and three synthetic Price distributions from the CART synthesized and the Bayesian synthesized datasets respectively. The CART synthesized Price data better follows the distribution of the confidential Price variable, while the Bayesian model synthesized Price data generally follows the distribution of the confidential Price variable but fails to capture some local features in the right tail of the confidential Price variable distribution.

\subsubsection*{Supp 4.1.1: Global Data Utility}
Similar to the global data utility evaluation process we took in the main text, we consider three utility measures: 1) the propensity score measure, 2) the cluster analysis measure, and 3) the empirical CDF measure \citep{woo2009global, Snoke2018JRSSA}. We evaluate the above-mentioned utilty measures for the CART-synthesized dataset and compare them with the utility results we obtained for the Bayesian model in the main text. All results are based on $m = 20$ synthetic datasets. Table \ref{tab:CART Data Utility} summarizes the data utility results. Both synthesis methods show high level of global utility. In comparison, CART synthesized data has better global data utility, which agrees with our visual inspect of the AvailableDays histogram and Price density plot. It is noteworthy that although we observe a ten fold difference in some of the utility measure scores between the CART and Bayesian synthesized datasets, this does not indicate the CART-synthesized dataset has ten times better data utility but only suggests that the CART-synthesized dataset has better utility than the Bayesian synthesized dataset. 

\begin{table}[H]
\centering
\begin{tabular}{>{\centering}p{1.6in} >{\centering}p{1in} >{\centering}p{0.8 in} >{\centering}p{1.1in} >{\centering}p{1in}}
\hline
  & Propensity $U_p$  & Cluster $U_c$ & eCDF-max $U_m$ & eCDF-avg $U_a$ \tabularnewline
\hline
AvailableDays\_Bayes & \multirow{2}{*}{$1.36\cdot 10^{-4}$} & \multirow{2}{*}{$4.63\cdot 10^{-5}$}  & 0.0314  & $2.82 \cdot 10^{-4}$  \tabularnewline
Price\_Bayes      & &  & 0.0509 & $3.94 \cdot 10^{-4}$ \tabularnewline
\hline 
AvailableDays\_CART & \multirow{2}{*}{$1.34\cdot 10^{-5}$}  & \multirow{2}{*}{$4.47\cdot 10^{-6}$} & $0.0083$  & $2.18\cdot 10^{-5}$\tabularnewline
Price\_CART &  &  & 0.0078 & $1.76\cdot 10^{-5}$\tabularnewline
\hline
High Utility & $\approx 0$  & $\approx 0$ & $\approx 0$  & $\approx 0$\tabularnewline
Low Utility & $\approx 1/4$  & large $U_c$ & large $U_m$ & large $U_a$\tabularnewline
\hline
\end{tabular}
\vspace{1mm}
\caption{Comparison of global utility measures between the Bayesian and CART synthesis.}
\label{tab:CART Data Utility}
\end{table}

\subsubsection*{Supp 4.1.2: Analysis Specific Utility}
In the main text, we evaluated the point estimate, the 95\% confidence interval, and the interval overlap measure of the AvailableDays and Price variables for key properties including the mean, the 25\% quantile, and the 90\% quantile. We apply the same evaluation process to the CART-synthesized datasets and compare the results with the Bayesian model results in the main text. Figure \ref{fig:mean_CI}, \ref{fig:Q25_CI}, and \ref{fig:Q90_CI} show the evaluation results of the point estimate, the 95\% confidence interval, and the interval overlap measure if the mean, the 25\% quantile, and 90\% quantile respectively. In each panel, we include the point estimate (box with a cross) and the 95\% confidence interval (the vertical bars) of the confidential data (black and left), of the Bayesian synthesized data (yellow and middle), and of the CART data (blue and right). Moreover, we add the interval overlap measure $I$ for each quantity in the panel. 

\begin{figure}[htb!]
    \centering
    \includegraphics[scale = 0.095]{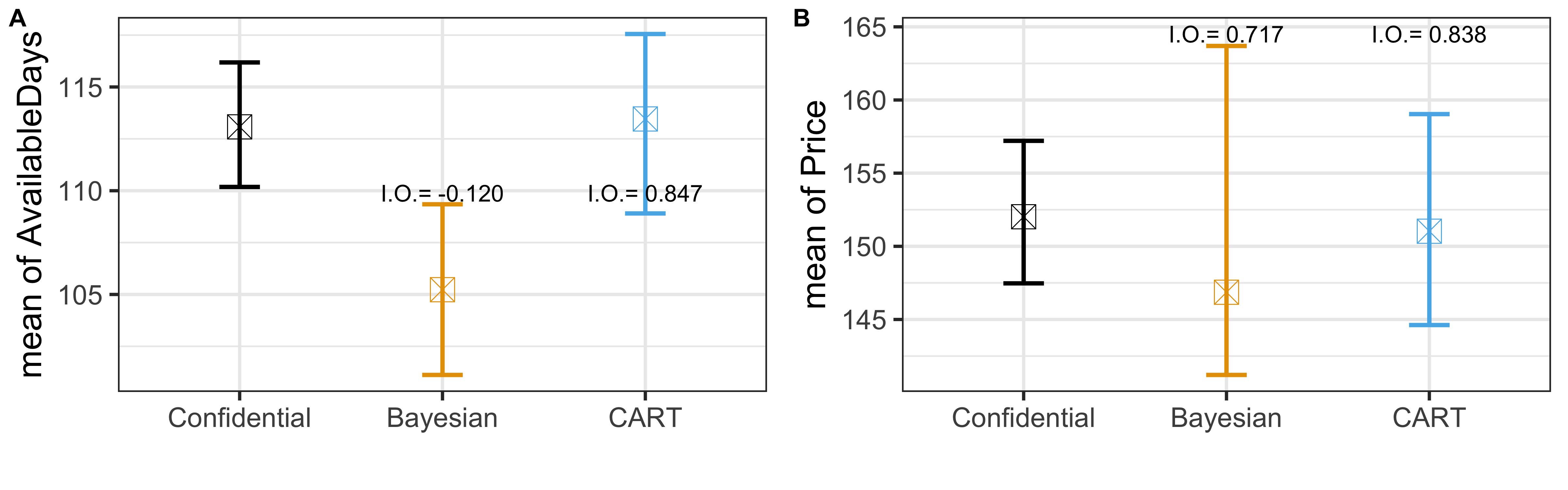}
    \caption{The point estimate, $95\%$ confidence interval, and interval overlap of the means of the confidential, the Bayesian synthesized, and the CART synthesized datasets. Panel \textbf{A} shows the estimand for AvailableDays and panel \textbf{B} shows the estimand for Price. I.O. stands for the interval overlap measure $I$.}
    \label{fig:mean_CI}
\end{figure}

\begin{figure}[htb!]
    \centering
    \includegraphics[scale = 0.095]{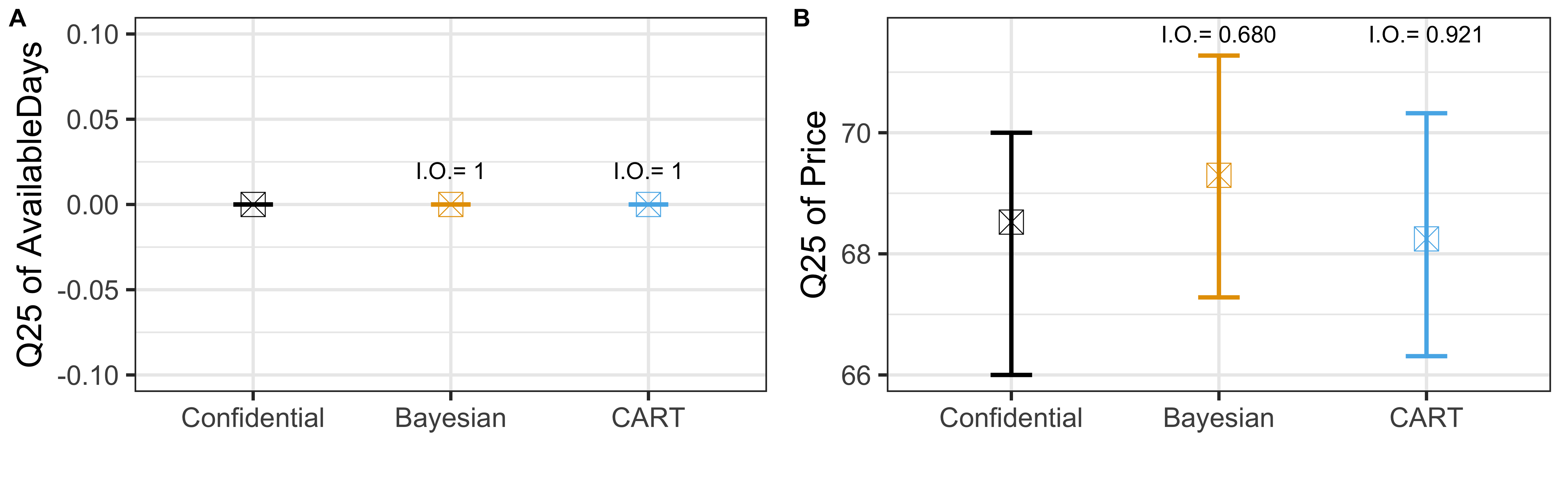}
    \caption{The point estimate, $95\%$ confidence interval, and interval overlap of the $25\%$ quantiles of the confidential, the Bayesian synthesized, and the CART synthesized datasets.  Panel \textbf{A} shows the estimand for AvailableDays and panel \textbf{B} shows the estimand for Price. I.O. stands for the interval overlap measure $I$.}
    \label{fig:Q25_CI}
\end{figure}

\begin{figure}[htb!]
    \centering
    \includegraphics[scale = 0.095]{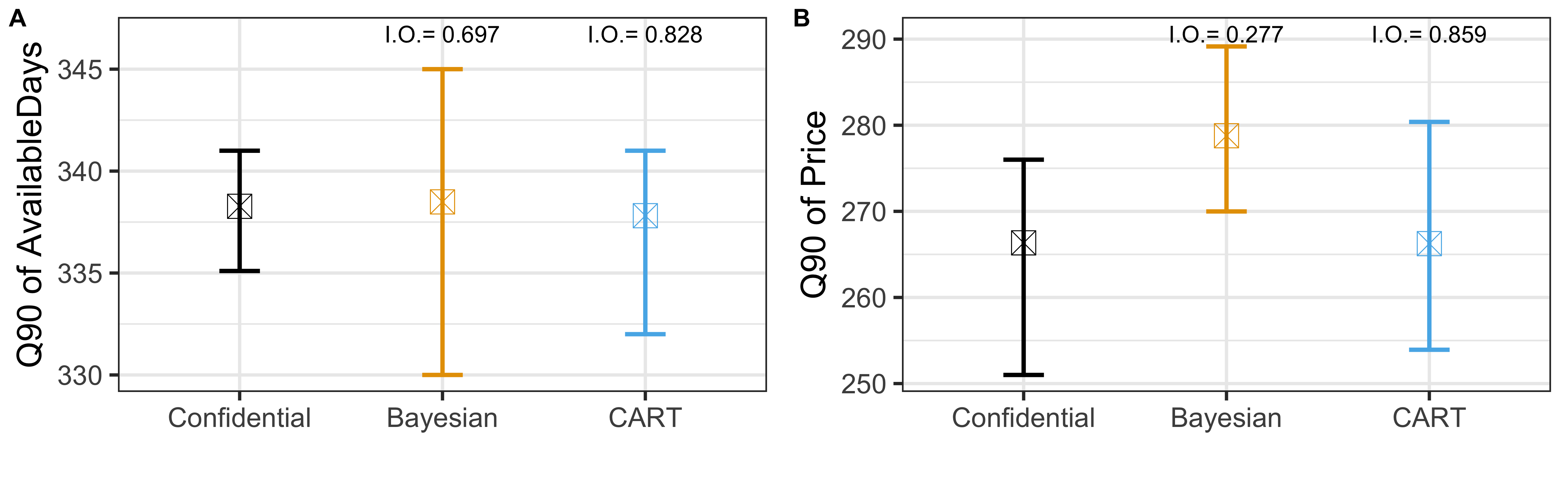}
    \caption{The point estimate, $95\%$ confidence interval, and interval overlap of the $90\%$ quantiles of the confidential, the Bayesian synthesized, and the CART synthesized datasets.  Panel \textbf{A} shows the estimand for AvailableDays and panel \textbf{B} shows the estimand for Price. I.O. stands for the interval overlap measure $I$.}
    \label{fig:Q90_CI}
\end{figure}

Overall, the point estimates of the 25\% and 90\% quantile of interest are close between the confidential data and the Bayesian synthesized data for AvailableDays and between the confidential data and the CART data, and the 95\% confidence interval overlap measures are close to 1 as well, suggesting high level of utility preservation. However, compared with the Bayesian synthesized data, the CART synthesized data perform better at preserving the mean, the 25\% quantile, and the 90\% quantile of the confidential dataset.

We also consider regression analysis types of analysis-specific utility measures as we did in the main text. We examine the regression coefficients of the effects of Neighborhood, RoomType, ReviewsCount, and AvailableDays on the response variable, Price. Applying the same regression coefficient estimation methodology as in the main text, we calculate the point estimate, the 95\% confidence interval, and the interval overlap measure for each regression coefficient. We compare the results among the confidential, the Bayesian synthesized, and the CART synthesized datasets. Selected results are shown in Figure \ref{fig:regression_CI}.

\begin{figure}[htb!]
    \centering
    \includegraphics[scale = 0.095]{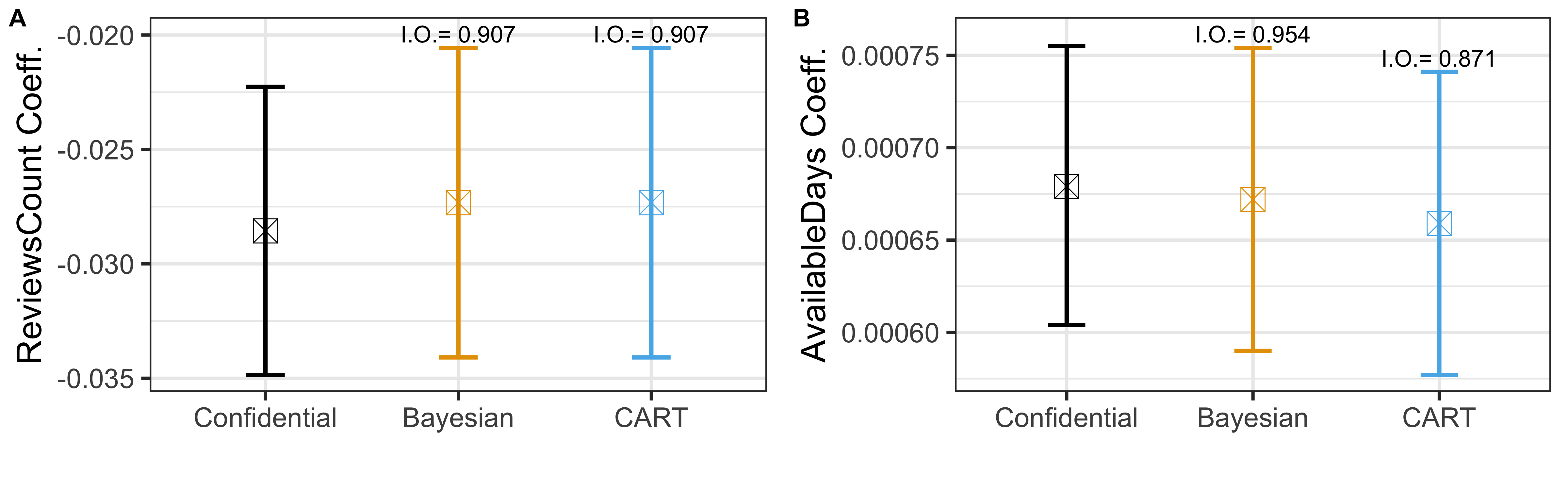}
    \caption{The point estimate, $95\%$ confidence interval, and interval overlap of selected regression coefficients from a linear regression analysis of the confidential, the Bayesian synthesized, and the CART synthesized datasets. Panel \textbf{A} shows the ReviewsCount coefficient and panel \textbf{B} shows the AvailableDays coefficient.}
    \label{fig:regression_CI}
\end{figure}

The point estimates of the coefficients for ReviewsCount and AvailableDays given the Bayesian synthesized data or the CART synthesized data are both close to those given the confidential data. In both cases, the interval overlap measures for both synthetic models are almost 1, indicating extremely high level of utility. There is no significant difference between the interval overlap measures between the Bayesian synthesized data and the CART synthesized data, though one might argue slightly better utility results of the Bayesian model, especially for the AvailableDays coefficient.

\subsection*{Supp 4.2: Disclosure Risk Results and Comparisons}

\subsubsection*{Supp 4.2.1: Attribute Disclosure Risk}

We adopt the same attribute disclosure risk method as we used in the main text to evaluate the attribute disclosure risk of the CART-synthesized dataset. Table \ref{tab:AR Compare} shows the comparison of attribute disclosure risk between the CART-synthesized dataset and the Bayesian synthesized dataset. Both synthesis models have a reduced level of attribute disclosure risk when compared with the inherent attribute disclosure risk of the confidential dataset. However, the Bayesian model offers better attribute disclosure risk protection than CART synthesis. Furthermore, since we report the attribute disclosure risk of a dataset as the sum of the attribute disclosure risk of all records in the dataset, the higher attribute disclosure risk of the CART-synthesized dataset suggests that on average, each record in the CART-synthesized dataset has a 1.5 to 2 times higher probability of exposing its AvailableDays and Price attributes to a data intruder.

\begin{table}[h]
\centering
\begin{tabular}{>{\centering}p{0.5in} >{\centering}p{0.5in} >{\centering}p{1.4 in} >{\centering}p{1.7 in} >{\centering}p{1.4 in}}
\hline
r\_avail & r\_price & Confidential AR & Bayesian AR & CART AR \tabularnewline
\hline
5 & 5\% & 636.40 & 126.07 & 238.53 \tabularnewline
\hline
10 & 5\% & 657.53 & 146.44 & 264.46 \tabularnewline
\hline
10 & 10\% & 816.55 & 291.73 & 433.51  \tabularnewline
\hline
\end{tabular}
\vspace{1mm}
\caption{Comparison of attribute disclosure risk between the confidential, the Bayesian synthesized, and the CART synthesized datasets.}
\label{tab:AR Compare}
\end{table}

\subsubsection*{Supp 4.2.2: Identification Disclosure Risk}
Using the same identification disclosure risk evaluation measures as we did in the main text, we evaluate the identification disclosure risk of the CART-synthesized dataset and compare the results with the inherent identification disclosure risk of the confidential dataset and the identification disclosure risk of the Bayesian synthesized dataset. Table \ref{tab:IR Compare} shows the resulting identification disclosure risk measures for all three dataset. Both synthetic datasets achieved significant reduction in identification disclosure risk compared to the confidential dataset. The CART-synthesized and the Bayesian synthesized datasets have comparable identification disclosure risk measures.

\begin{table}[h]
\centering
\begin{tabular}{>{\centering}p{1.5in} >{\centering}p{0.9in} >{\centering}p{1in} >{\centering}p{1in} >{\centering}p{1in}}
\hline
& EMR & TMR & FMR & u \tabularnewline
\hline
Confidential IR & 7182.03 & 0.623 & 0 & 6226 \tabularnewline
\hline
Bayesian IR & 125.59 & 0.010 & 0.912 & 1153 \tabularnewline
\hline
CART IR & 111.54 & 0.007 & 0.951 & 1367 \tabularnewline
\hline
\end{tabular}
\vspace{1mm}
\caption{Comparison of identification disclosure risk between the confidential, the Bayesian synthesized, and the CART synthesized datasets.}
\label{tab:IR Compare}
\end{table}

\subsection*{Supp 4.3: Summary}

Overall, CART offers better global data utility and analysis-specific estimates for the mean, the 25\% quantile, and the 90\% quantile, and has comparable analysis-specific estimates for regression coefficients and identification disclosure risk. Our Bayesian zero-inflated Poisson model has lower attribute disclosure risks.

\end{document}